\pdfoutput=1 
\documentclass[12pt]{article}

\usepackage{graphicx}
\usepackage{amsmath}
\usepackage{amssymb}
\usepackage{lineno}
\usepackage{url}
\usepackage{xspace,multicol}
\usepackage{siunitx}
\usepackage{subcaption}
\usepackage{color, colortbl}
\usepackage{units}
\usepackage{ragged2e}
\usepackage{array}
\usepackage{tabularx}
\usepackage{authblk}
\usepackage{heppennames}
\usepackage{feynmp}
\usepackage{libertine}
\usepackage{textgreek}
\newcommand{\guineapig}{GuineaPig\xspace}

\newcommand{\murm}{%
  \ifmmode
    \mathchoice
        {\hbox{\normalsize\textmu}}
        {\hbox{\normalsize\textmu}}
        {\hbox{\scriptsize\textmu}}
        {\hbox{\tiny\textmu}}%
  \else
    \textmu
  \fi
}

\usepackage{authblk}
\usepackage{fixltx2e}

\begin{document}


\title{Pair Background Envelopes\\in the SiD Detector\vspace*{0.3cm}\\{\normalsize Talk presented at the\\International Workshop on Future Linear Colliders (LCWS2016)\\Morioka, Japan, 5-9 December 2016. C16-12-05.4.}}

\author[1,2]{Anne Sch\"utz}

\affil[1]{\normalsize Karlsruhe Institute of Technology (KIT), Department of Physics, Institute of Experimental Nuclear Physics (IEKP), Wolfgang-Gaede-Str. 1, 76131 Karlsruhe}
\affil[2]{\normalsize Deutsches Elektronen-Synchrotron (DESY), Notkestr. 85, 22607 Hamburg}

\maketitle


\begin{abstract}
The beams at the ILC produce electron positron pairs due to beam-beam interactions.
This note presents for the first time a study of these processes in a detailed simulation, which shows that these pair background particles appear at angles that extend to the inner layers of the detector.
The full data set of pairs produced in one bunch crossing was used to calculate the helix tracks, which the particles form in the solenoid field of the SiD detector.
The results suggest to further study the reduction of the beam pipe radius and therefore to either add another SiD vertex detector layer, or reduce the radius of the existing vertex detector layers, without increasing the detector occupancy significantly.
This has to go along with additional studies whether the improvement in physics reconstruction methods, like c-tagging, is worth the increased background level at smaller radii.
\end{abstract}

\section{Introduction}
\label{sec:introduction}
Since the pair background scatters on material and irradiates it over time, the ideal distance between the area of high density pair background and the beam pipe can be decided by studying the envelopes of the pair background helixes in the magnetic solenoid field.
Takashi Maruyama has previously done this study with different beam parameters~\cite{Takashi_plot}, but with applying cuts to the data set because of limited CPU power~\cite{Takashi}.
The plots shown in this proceeding, in contrast, were done with a full pair background data set from the most recent simulations without any cuts applied.\\
With the beam pipe radius of \unit{1.2}\,{cm} in the immediate interaction point region and the beam pipe increasing in radius in a cone shape, the fraction of pairs leaving the beam pipe can be calculated in order to convey an understanding of how many pair background particles enter the SiD detector and interact with the material. 

\section{The pair background envelopes}
\label{Detector}
\begin{figure}
    \centering
    \includegraphics[width=0.4\textwidth]{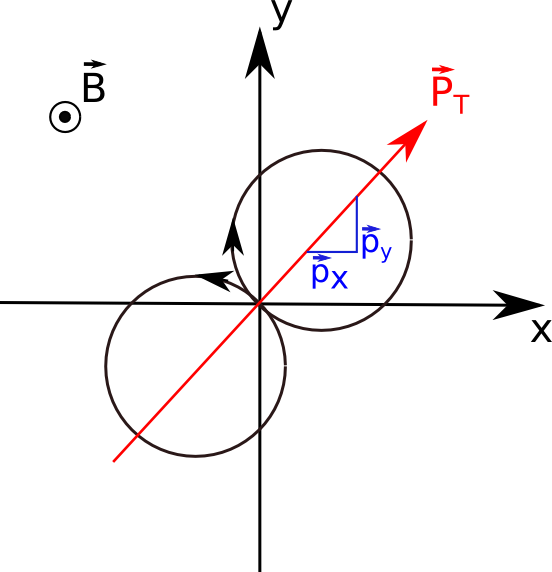}
    \caption[Projection of the helix on the xy-plane]{
    This schematic shows the vector of the transverse momentum (P\textsubscript{T}) in dependency of the x- and y-momenta (p\textsubscript{x} and p\textsubscript{y}).
    With a constant magnetic field applied, a charged particle performs a helix track which is projected into a circle on the displayed xy-plane.
    Depending on the charge of the particle, the direction of the rotation is either clockwise or anticlockwise.
    The center and orientation of the projected circle is dependent on the transverse momentum of the particle.
    }
    \label{fig:helix_circle}
\end{figure}

\begin{figure}
    \centering
    \begin{subfigure}[b]{0.49\textwidth}
    \centering
        \includegraphics[height=0.26\textheight]{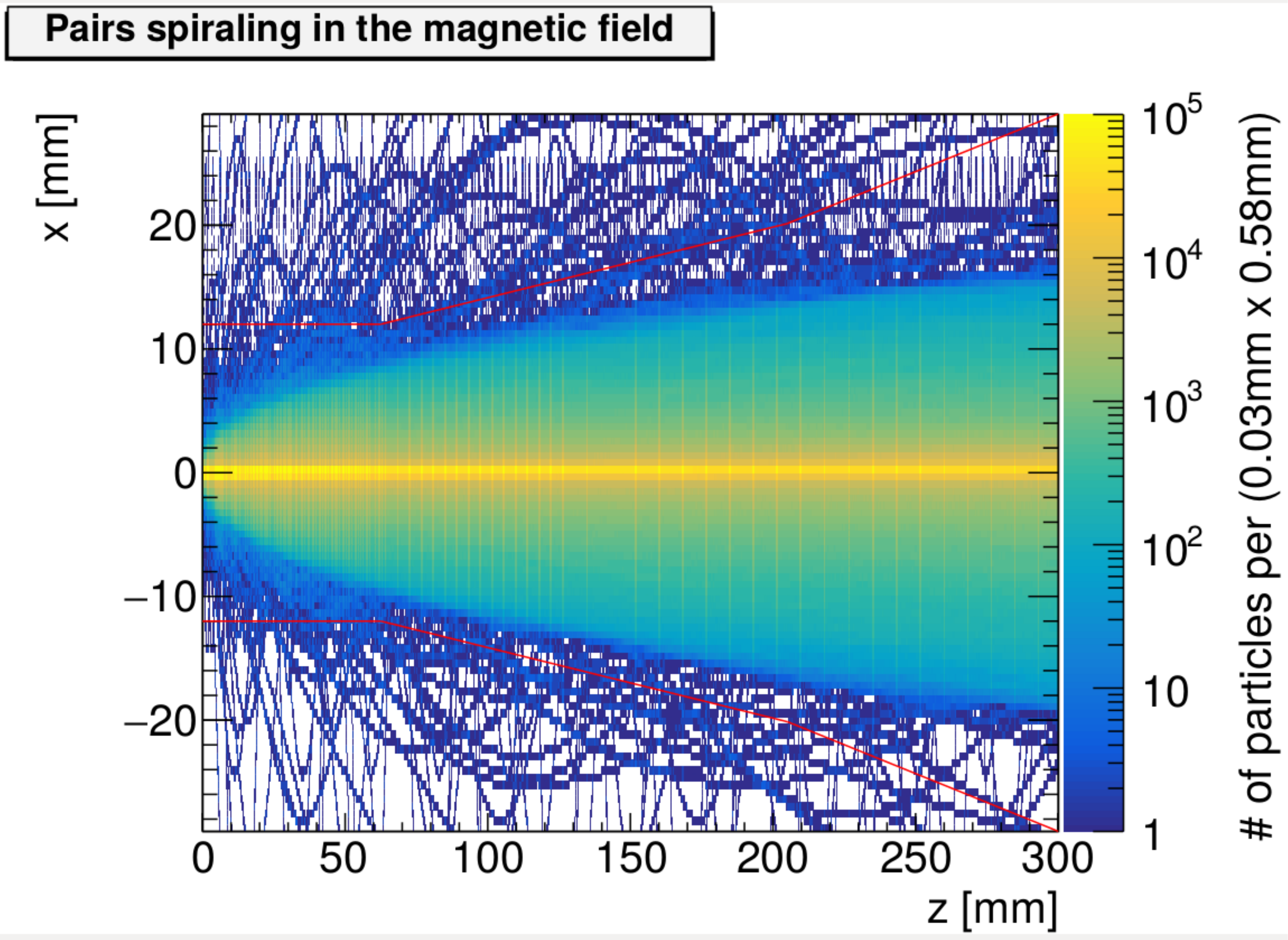}
        \caption{Helix tracks projected on the xz plane}
	\label{fig:helix_xz}
    \end{subfigure}
    \begin{subfigure}[b]{0.49\textwidth}
    \centering
        \includegraphics[height=0.26\textheight]{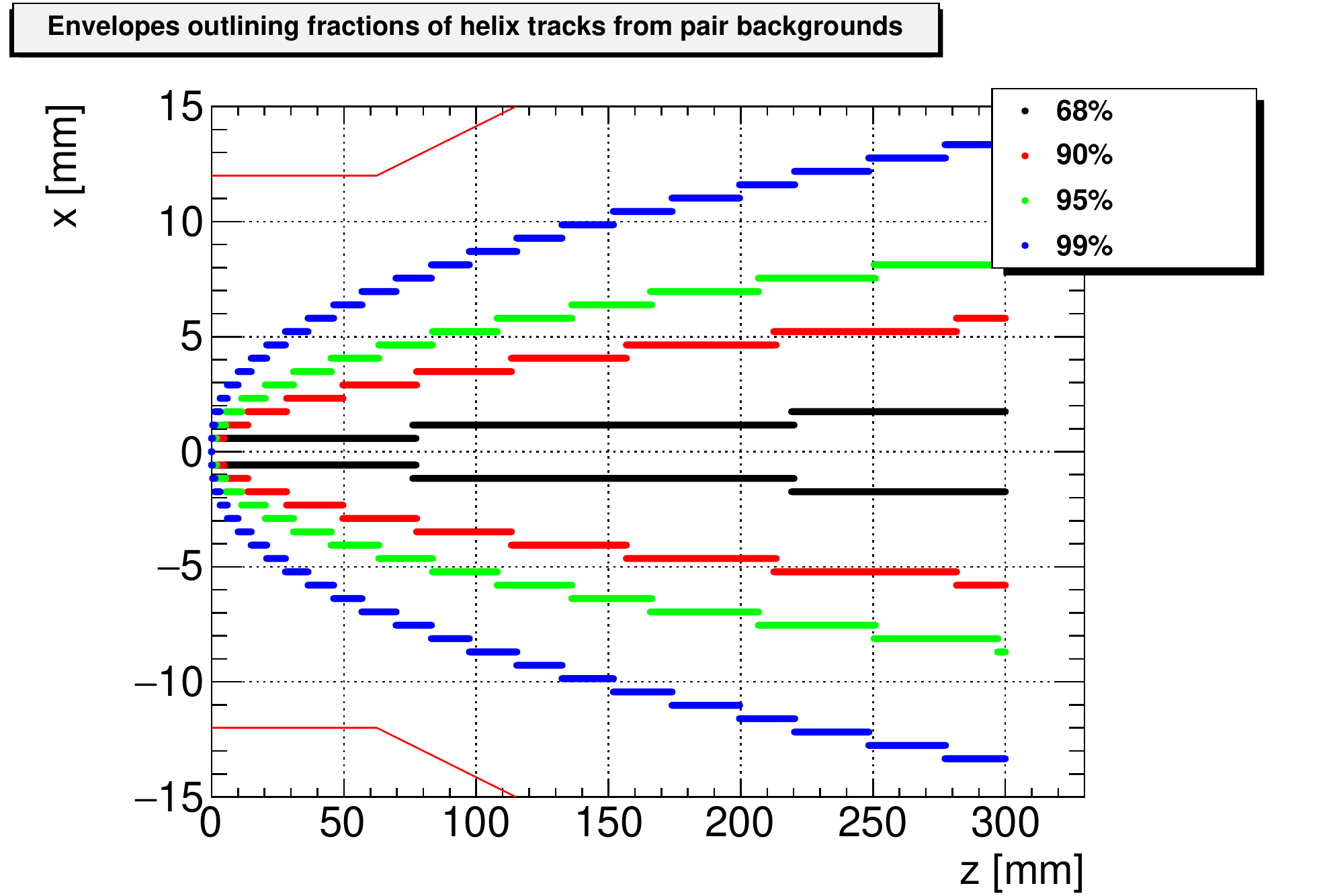}
        \caption{Envelopes on the xz plane}
        \label{fig:envelopes_xz}
    \end{subfigure}\\
    \begin{subfigure}[b]{0.49\textwidth}
    \centering
        \includegraphics[height=0.26\textheight]{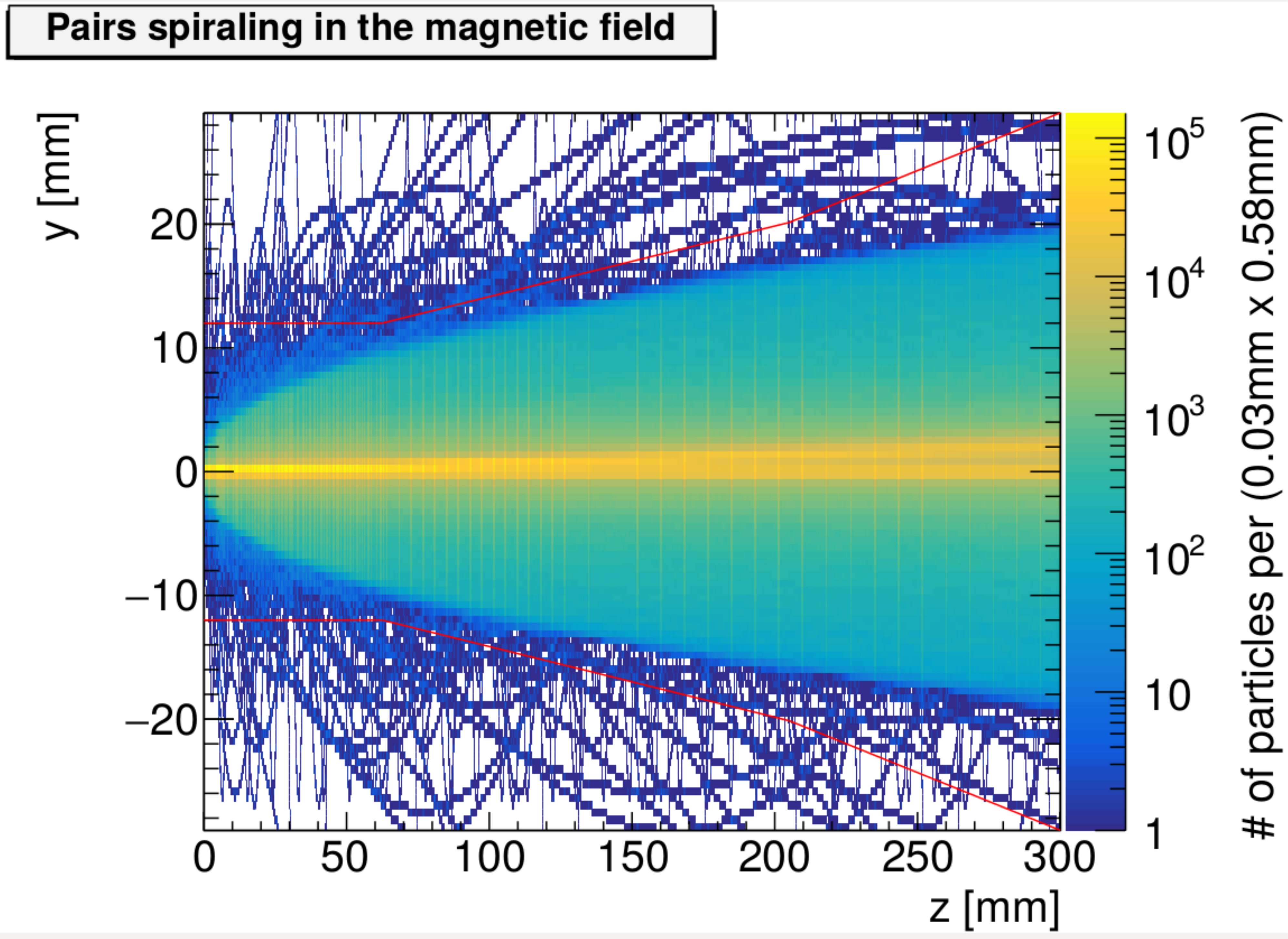}
        \caption{Helix tracks projected on the yz plane}
	\label{fig:helix_yz}
    \end{subfigure}
    \begin{subfigure}[b]{0.49\textwidth}
    \centering
        \includegraphics[height=0.26\textheight]{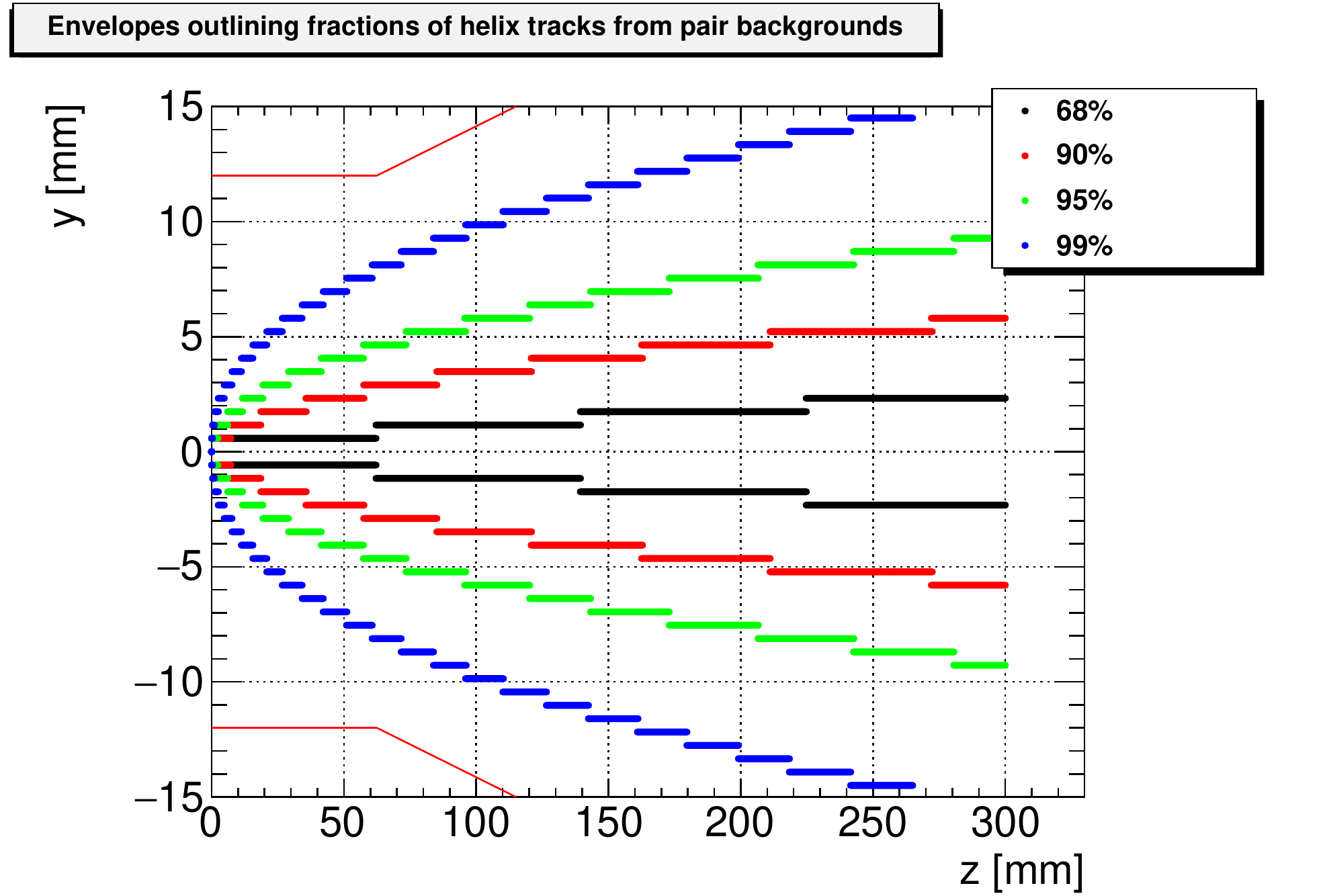}
        \caption{Envelopes on the yz plane}
        \label{fig:envelopes_yz}
    \end{subfigure}
    \caption[Helix tracks and their envelopes]{
    The projection of the helix tracks from the pair background particles of one bunch crossing are shown in the xz- and yz-plane in Figures a) and c).
    The color scales shows how many particle tracks are in the single bins of these plots.
    To get a better grasp, Figures b) and d) show the envelopes outlining certain fractions of helix tracks.
    Therefore, the blue line represents the envelope of 99\% of all pair tracks.
    In all subfigures, the thin red lines represent the beam pipe.
    The helix tracks were calculated for a homogeneous magnetic field of \unit{5}\,{T}.
    }
    \label{fig:Helixes}
\end{figure}

\begin{figure}
    \centering
    \begin{subfigure}[b]{0.49\textwidth}
    \centering
        \includegraphics[height=0.21\textheight]{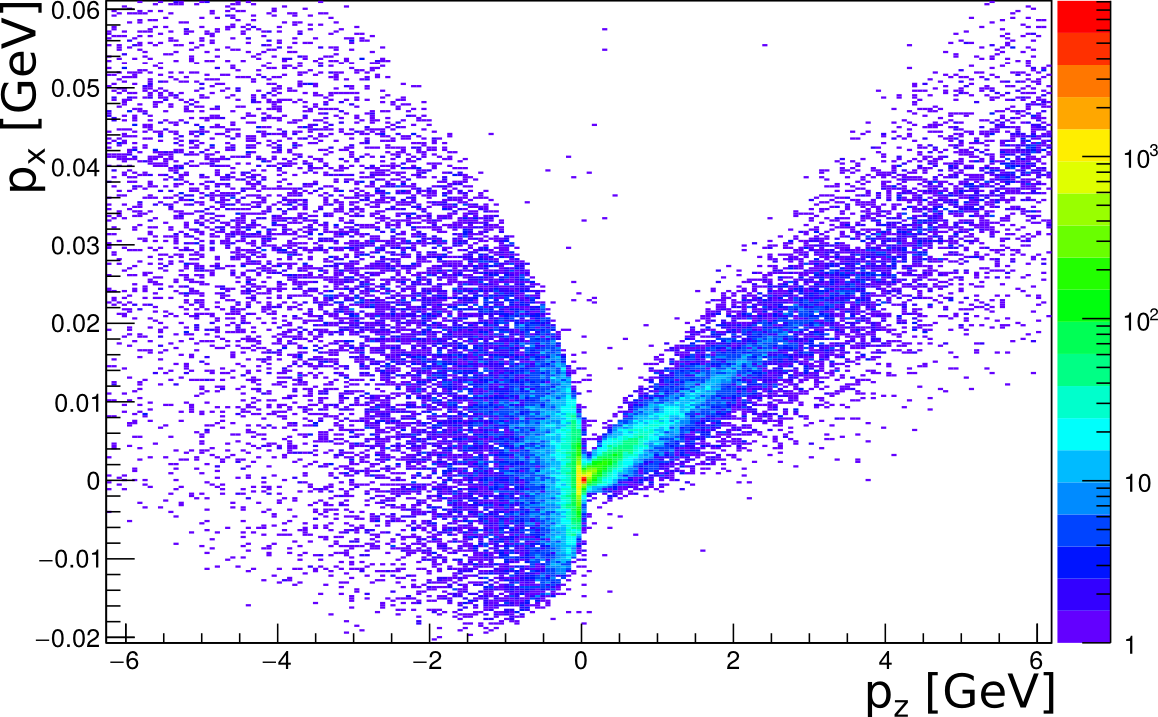}
        \caption{p\textsubscript{x} vs. p\textsubscript{z}, \Pem}
	\label{fig:px_pz_ele}
    \end{subfigure}
    \begin{subfigure}[b]{0.49\textwidth}
    \centering
        \includegraphics[height=0.21\textheight]{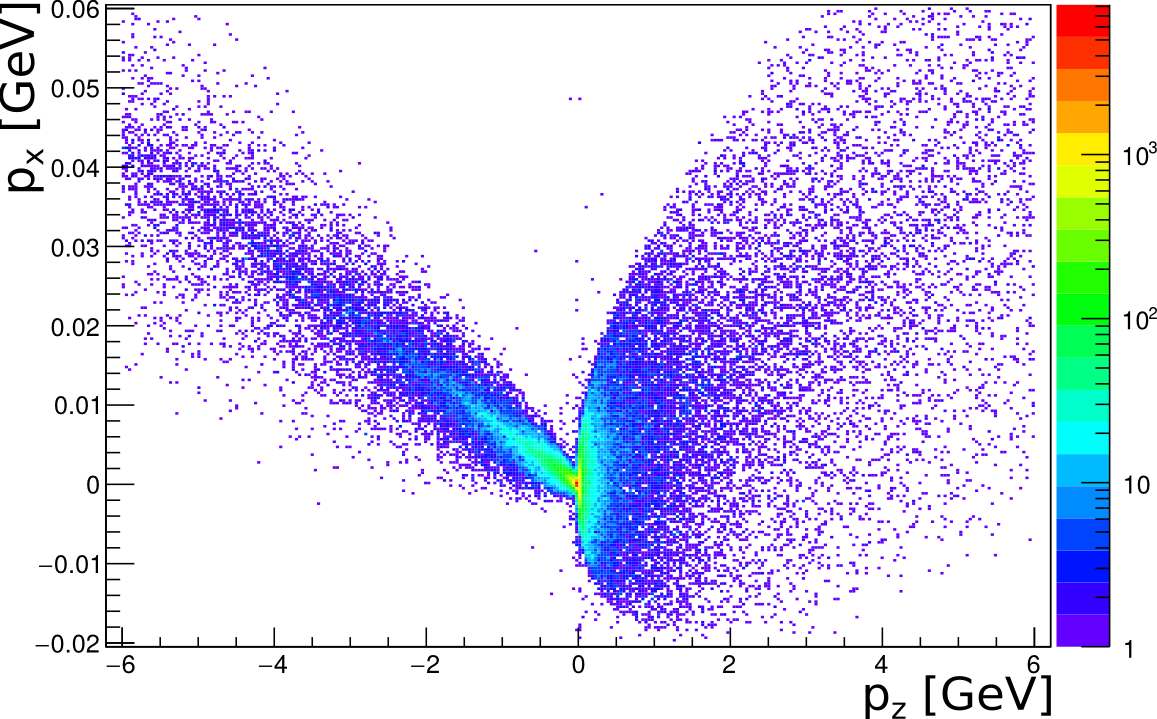}
        \caption{p\textsubscript{x} vs. p\textsubscript{z}, \Pep}
        \label{fig:px_pz_posi}
    \end{subfigure}\\
    \begin{subfigure}[b]{0.49\textwidth}
    \centering
        \includegraphics[height=0.21\textheight]{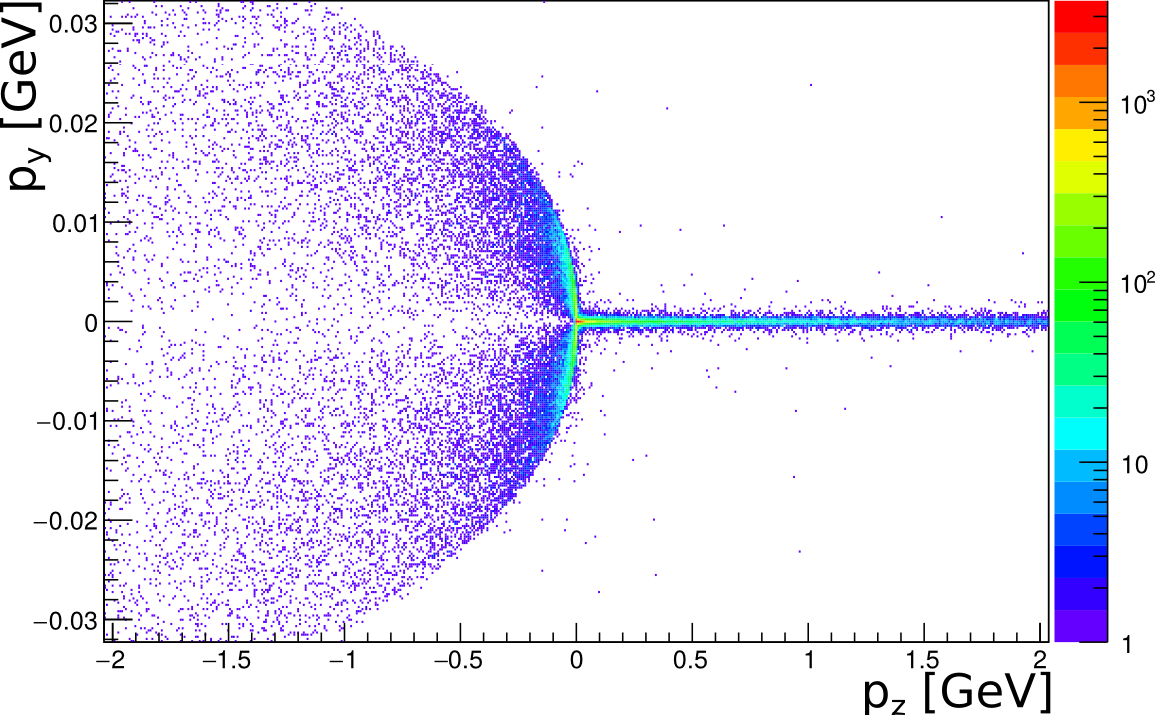}
        \caption{p\textsubscript{y} vs. p\textsubscript{z}, \Pem}
	\label{fig:py_pz_ele}
    \end{subfigure}
    \begin{subfigure}[b]{0.49\textwidth}
    \centering
        \includegraphics[height=0.21\textheight]{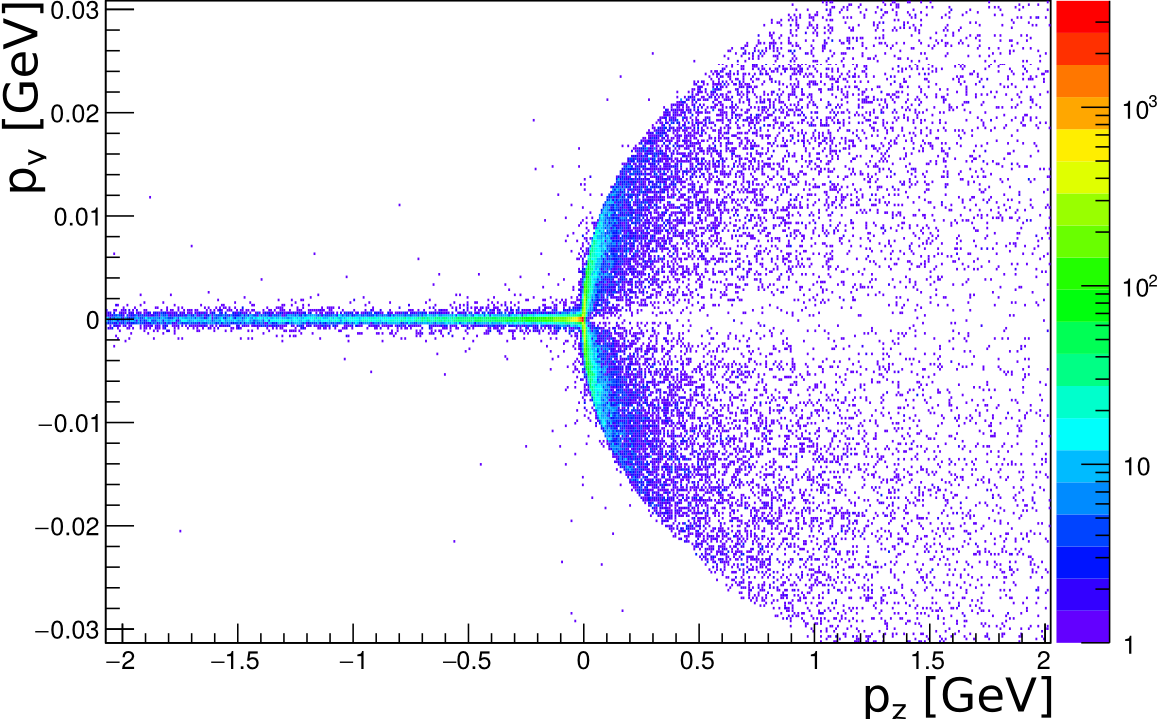}
        \caption{p\textsubscript{y} vs. p\textsubscript{z}, \Pep}
        \label{fig:py_pz_posi}
    \end{subfigure}
    \caption[Momentum distributions for electrons and positrons]{
    The pair background particle momenta are plotted for electrons and positrons separately for all pairs from one bunch crossing.
    It becomes clear that the particle momenta distributions are quite distinctive, and that the distributions are mirrored for electrons and positrons.
    }
    \label{fig:momenta}
\end{figure}

The pair background induced by beam-beam interactions, in which the beamstrahlung from the initial beams produces secondary \Pep \Pem pairs, is simulated with \guineapig, a Monte Carlo (MC) background event generator~\cite{Schulte:1997nga}.
For this study, the pairs are generated for the ILC-500GeV beam parameters.
The 4-vectors of the resulting pair background particles from one bunch crossing are then the input to a helix algorithm, calculating the tracks of the pairs in the SiD detector.\\
The algorithm takes the particle vertices, their momenta and charge as arguments, and calculates the radius of the helix, its center position and its pitch for a given homogeneous magnetic field.
For this, the following assumptions are made: 
The particle momenta does not change in the region shown in the plots, because of which the helix radius is constant over this distance.
Additionally, any particle interaction with other particles or the material is not taken into account.
Since the study was done for the pair background in the SiD detector, the helic track calculations were done for a solenoid field strength of \unit{5}\,{T}.
Figure~\ref{fig:helix_circle} shows schematically the projection of a helix onto the xy-plane with the transverse momentum (P\textsubscript{T}) and the x- and y-momenta (p\textsubscript{x} and p\textsubscript{y}) of the particle being displayed, on which the helix is dependent.\\
Figure~\ref{fig:Helixes} shows the resulting helix tracks of the pair background from one bunch crossing in the xz- and yz-plane.
With the envelopes drawn in Figures~\ref{fig:envelopes_xz} and \ref{fig:envelopes_yz}, it becomes clear that the distance between 99\% of all pair particle tracks and the beam pipe is more than \unit{4}\,{mm} at any given point.
Overall, with the current beam pipe design, only $\sim$ 0.45\% of all pairs leave tracks outside the beam pipe.\\
With a closer look, one can see that the envelopes are broader in the yz-plane than in the xz-plane, and that the distribution of tracks in the yz-plane is asymmetric.
The reason for that are the initial momentum distributions of the pair background particles: the x-momentum distributions are broader than for the y-momentum, because of which the projected helix circles on the xy-plane are orientated more around the x-axis.
Hence, the distribution of the projection of all helix tracks on the yz-plane is wider than on the xz-plane.\\
The asymmetry of the helix track distribution in y is due to the different momenta of positrons and electrons in the beam direction, which can be seen in Figure~\ref{fig:momenta}.
The secondary particles are boosted in the direction of the initial beam with the same charge, i.e. the secondary electrons are mainly boosted in the direction of the \Pem beam, and the secondary positrons in the other direction accordingly.
Looking at both electrons and positrons, the distribution of track density becomes clear.

\section{Summary, Conclusions, and Outlook}
The pair background envelopes in the SiD detector were studied by using the up-to-date \guineapig data set of one bunch crossing for the ILC500GeV parameters, and taking the pair particle 4-vector as input to a Helix algorithm that calculates the helix tracks under certain assumptions.
The presented results from this study suggest that the ILC community could think about reducing the beam pipe radius by at least \unit{2}\,{mm} without exposing material to the pair background.
The SiD group would then have the opportunity to either add another vertex detector layer with a smaller radius or to shrink the radius of the existing vertex detector layers.
Whilst this opens the possibility for improvements in physics event reconstruction through c-tagging for example, further studies of the level of background and synchrotron radiation at such
radii need to be performed.

\section*{Acknowledgments}
The author would like to thank Takashi Murayama (SLAC) for useful discussions and clarifications of his work in \cite{Takashi_plot}.



\bibliographystyle{unsrt}
\bibliography{bibliography.bib}

\begin{thebibliography}{1}

\bibitem{Takashi_plot}
Takashi Maruyama.
\newblock {Background studies, GDE Baseline Assessment Workshop, SLAC}.
\newblock
  \url{http://agenda.linearcollider.org/event/4612/contributions/18939/attachments/15265/24970/BAW-20110119.pdf},
  Jan 2011.

\bibitem{Takashi}
Takashi Maruyama.
\newblock {Background studies}.
\newblock {Private email conversation}.

\bibitem{Schulte:1997nga}
Daniel Schulte.
\newblock {\em {Study of Electromagnetic and Hadronic Background in the
  Interaction Region of the TESLA Collider}}.
\newblock PhD thesis, DESY, 1997.

\end{thebibliography}

\end{document}